\documentstyle[12pt,epsf]{article}
\newcommand{\LL}{\Lambda\Lambda}

\def\tstrut{\vrule height2.5ex depth0pt width0pt} % used in tables
\newcommand{\be}{\begin{equation}}
\newcommand{\bea}{\begin{eqnarray}}
\newcommand{\ee}{\end{equation}}
\newcommand{\eea}{\end{eqnarray}}

\begin{document}
\begin{titlepage}

  \vspace*{5mm}

\vspace*{2cm}

\begin{center}
{\Large \bf Pionic decay of $\Lambda$ Hypernuclei in a Continuum
Shell Model}

\vspace{1.5cm}
{\large{\bf C. Albertus, J.E. Amaro and J. Nieves}}\\[2em]
Departamento de F\'{\i}sica Moderna, Universidad de Granada, 
E-18071 Granada, Spain.

\end{center}

\vspace{2cm}
\begin{abstract}
We evaluate pionic decay widths of $\Lambda$ hypernuclei using a 
shell model for, both  the nuclear bound  and the continuum
nucleon wave functions in the final state, and distorted 
waves for the outgoing pion. An excellent agreement with the recent KEK 
measurement of $\pi^-$-decay widths of $^{12}_{\phantom{a}\Lambda}{\rm C}$
and $^{28}_{\phantom{a}\Lambda}{\rm Si}$ is found. Besides, results for 
 $^{56}_{\phantom{a}\Lambda}{\rm Fe}$ are consistent with the existing
upper bound.

\vspace{1cm}

\noindent
{\it PACS: 21.80+a, 23.90.+w, 25.80.Pw,13.30.-a, 23.20.Js, 21.60.Cs}  \\
{\it Keywords: $\Lambda$ hypernuclei, weak decay, mesonic decay, continuum
shell model} . 
\end{abstract}

\end{titlepage}

\newpage

\setcounter{page}{1}

\section{Introduction}

The pionic decay of $\Lambda$ hypernuclei has received attention in
the past both from the experimental~\cite{Mo74}--\cite{Sa01} and
theoretical~\cite{Da58}--\cite{Al00} points of view (for a recent
review see Ref.~\cite{AG02}). It is well known that the mesonic decay
is largely suppressed by Pauli blocking, although the consideration of
proper pion distorted waves~\cite{IMB88,NO92,Mo94}
 weakens the effect of the Pauli blocking
considerably. On the other hand, it was also pointed out in
Ref.~\cite{NO92} the relevance of the use of a correct energy balance
in the decay. Two theoretical frameworks have been traditionally used
to describe this process: Polarization Propagator Method
(PPM)~\cite{OS85,ROS94,ZP99}, supplemented by quantum
field theory functional techniques~\cite{Al00,AG02}, and the
Wave Function Method (WFM)~\cite{IMB88}--\cite{St93}. In the PPM, the
calculation of the pionic widths is performed in nuclear matter and
the results are translated to finite nuclei by means of the Local
Density Approximation (LDA). Both methods PPM and WFM turn out to be
related in a semiclassical limit~\cite{O94}, being the WFM more
reliable than the PPM+LDA, since the mesonic decay channel is quite
sensitive to the details of the shell structure of the hypernuclei,
specially for light systems.

In Ref.~\cite{NO92} a simple  Woods-Saxon (WS) shell model was employed, 
with a central potential of constant depth for several nuclei. This
global potential did not include 
spin-orbit term and only the radius of the WS well
depended on the specific nuclei ($ \propto A^\frac13 $). The
resulting  binding energies of the shells were globally shifted, by hand,
to reproduce the ground state masses of the involved nuclei. Besides,
the continuum contribution to the decay was estimated by discretizing
the positive nucleon energy levels by means of an infinite barrier
placed at distances of about 20 fm. Despite  all these crude
approximations, the model  led to  predictions for
the mesonic decay widths of $^{12}_{\phantom{a}\Lambda}{\rm C}$
which were compatible, within errors, with the available experimental
measurements at that time, for both $\pi^0$ and $\pi^-$ decay
channels. Furthermore, the model of Ref.~\cite{NO92} provided
 an overall description of medium and heavy hypernuclei mesonic
decay, which is still nowadays accepted as one of the most reliable
theoretical estimates~\cite{AG02}. 

Very recently, precise measurements of the $\pi^-$ mesonic decay of
$^{12}_{\phantom{a}\Lambda}{\rm C}$ and
$^{28}_{\phantom{a}\Lambda}{\rm Si}$ have been obtained at KEK, and an
upper bound for $^{56}_{\phantom{a}\Lambda}{\rm Fe}$ has also been
given~\cite{Sa01}. The purpose of this paper is to update the model of
Ref.~\cite{NO92}, improving the nuclear structure description of the
process, by using: i) more realistic potentials for each one of the
nuclei by fitting the parameters of a WS well, with 
spin-orbit forces, to  the ground state  and also to some of
the discrete final state energies of the nuclei involved in
the decay and ii) a Continuum Shell
Model (CSM) to describe the positive energy tail of the mesonic decay
width. In this way we replace the discrete sum used in
Ref.~\cite{NO92} by the appropriate integral over the continuum
states. This correct treatment of the continuum contribution is
important, since, as we shall see,  for
 $\pi^-$-$^{28}_{\phantom{a}\Lambda}{\rm Si}$ decay it amounts about
50\% of the total measured mesonic width, in contrast to the
case of the hypernuclei studied in
Refs.~\cite{IMB88} and~\cite{NO92}, for which the continuum contribution was 
  much smaller.

  Since, as mentioned above, the use of pion distorted waves turns out
  to be crucial~\cite{NO92}, \cite{Mo94}, the improved treatment of
  the nuclear structure allows us to use the new and precise
  measurements performed at KEK, as a further test of the quality of
  the pion-nucleus dynamics used in Ref.~\cite{NO92} and developed in
  the works of Refs.~\cite{NOG93} and \cite{NOG93bis}. A deep and
  detailed knowledge of the pion dynamics inside of a nuclear medium
  has become a topic of renewed interest to explore possible partial
  chiral restoration in the medium~\cite{wolfram}-\cite{KKW02}.

The paper is organized as follows: In Sect.~\ref{sec:mes} the needed
formulae to compute the pionic decay width are given, both for the discrete
and  continuum contributions. Details on the nuclear CSM used
in this work are discussed in Sect.~\ref{sec:ncs}, where the energy
reaction balance is also studied. Finally in Sect.~\ref{sec:con} we
present the results of this work and our main conclusions.

\section{Theoretical Description of the Pionic Decay.}
\label{sec:mes}

In this work, we compute the mesonic decay width as the sum of the 
contributions of the following processes\footnote{We use the notation 
$^{A}_{\Lambda}Z$, to denote  a nuclear core $^{A-1}Z$ and one bound
$\Lambda$ particle.}
\begin{eqnarray}
^{A}_{\Lambda}{\rm Z} & \to & \left(^{A}{\rm
Z}\right)_{d} +
\pi^0 , \label{eq:decay1}\\
^{A}_{\Lambda}{\rm Z} & \to & \left(^{A}{\rm
(Z+1)}\right)_{d} +
\pi^- ,\label{eq:decay2}\\
^{A}_{\Lambda}{\rm Z} & \to & \left(^{A-1}{\rm Z}
\right)_{gs} + n + 
\pi^0 , \label{eq:decay4}\\
^{A}_{\Lambda}{\rm Z} & \to & \left(^{A-1}{\rm Z}
\right)_{gs} + p + 
\pi^- ,\label{eq:decay3}
\end{eqnarray}
\noindent where $d$ denotes the ground or discrete excited states of the final
nucleus. As we shall see, we evaluate the processes of
Eqs.(\ref{eq:decay1}),~(\ref{eq:decay2})  by putting the outgoing
nucleon, coming from the weak $\Lambda$ decay, in an unoccupied bound
shell of the daughter nucleus. On the other hand, in
Eqs.~(\ref{eq:decay4}) and~(\ref{eq:decay3})   $gs$ means that the
daughter nucleus is left on its ground state.  In the two last
reactions the outgoing nucleon, coming from the weak $\Lambda$ decay, 
goes to the continuum (positive energy) and we denote that
contribution  by $\Gamma_c$, 
while the first two reactions give what we call discrete contribution
$\Gamma_d$. Thus we split the pionic width into two contributions, 
$\Gamma = \Gamma_c +
\Gamma_d$. Experimentally, what can be observed are the inclusive processes 
\begin{eqnarray}
^{A}_{\Lambda}{\rm Z} & \to & X + 
\pi^0 , 
\label{eq:decayq1}\\
^{A}_{\Lambda}{\rm Z} & \to &  X + \pi^- . 
\label{eq:decayq2}
\end{eqnarray}
The main contribution to these processes is given by the exclusive
ones shown in Eqs.~(\ref{eq:decay1}--\ref{eq:decay3}).

The pionic decay is produced by a baryonic one-body operator
\begin{equation}
\delta \widetilde{H}_{\Lambda \pi N}^\lambda = -G m_\pi^{2} \left 
\{S - \frac{P}{m_\pi} \,
\vec{\sigma} \cdot \vec{q}_{cm}\, \right \}\, \tau^{\lambda},\label{eq:decay-op}
\end{equation}
\noindent where $(G m_\pi^{2})^{2}/8 \pi = 1.945 \times 10^{-15}$, 
the constants $S$ and $P$ are equal to 1.06 and 0.527 respectively
and $m_\pi$ is the pion mass (139.57 or 134.98 MeV for $\pi^-$ or
$\pi^0$), $\vec{q}_{cm}$ is the momentum of 
the outgoing pion in the $\Lambda$ rest frame, and the 
Pauli matrices $\vec{\sigma}$ and $\tau^{\lambda}$, where $\lambda$ is
a cartesian isospin index which will be contracted to the pion field, 
act on the spin and
isospin Hilbert spaces respectively. Taking the $\Lambda$ isospin wave
function as that of a neutron, the $\tau^{\lambda}$ operator in
Eq.~(\ref{eq:decay-op}) implements the extreme $\Delta T = 1/2$ rule, which
leads to a rate of $\Lambda \rightarrow \pi^{-} p$  twice as large
as that of $\Lambda \rightarrow \pi^{0} n$.

The  free space $\Lambda$ decay width is readily evaluated and 
leads for proton or neutron
decay to
\begin{eqnarray}
\Gamma^{\,(\alpha)}_{free} & = & C^{\,(\alpha)} \, 
\frac{(G m_\pi^{2})^{2}}{4 \pi}
\frac{M\, q_{cm}}{M_{\Lambda}}\,\left\{ S^{2} +
\left(\frac{P}{m_\pi}\right)^{2} 
q^{2}_{cm}\right \}, \quad \alpha=p,n \label{eq:free-decay}\\
q_{cm} & = & \frac{\lambda^{1/2} (M^{2}_{\Lambda}, M^{2},
m_\pi^{2})}
{2M_{\Lambda}}, \\
\lambda (x,y,z) &=& x^2 + y^2 + z^2 -2xy-2xz-2yz,
\end{eqnarray}
where $\alpha$ indicates  neutron or proton or equivalently 
$\pi^0$ or $\pi^-$ decay, $C^{\,(p)} = 4$  and  
$C^{\,(n)} = 2$ are  isospin coefficients, 
and $M$ (938.27 or 939.57 MeV for $p$ or $n$) and 
$M_\Lambda$ (1115.68 MeV)  are the nucleon and $\Lambda$
masses respectively. The total free space hyperon
$\Lambda$ width, $\Gamma_\Lambda$, is given by the sum of proton
($\pi^-$) and neutron ($\pi^0$) contributions,
\begin{equation}
\Gamma_\Lambda = \Gamma^{\,(p)}_{free} + \Gamma^{\,(n)}_{free} \label{eq:free}
\end{equation}

In the case of a bound hypernucleus, assuming that the hyperon
$\Lambda$ is in the $1s_{\frac12}$ shell and a closed shell structure
for the underlying nuclear system, $^{A-1}{\rm Z}$, the width for any
of the processes of Eqs.~(\ref{eq:decay1}--\ref{eq:decay3}) is
given by\footnote{The following expressions are valid for the discrete
contribution (Eqs.~(\ref{eq:decay1}) and~(\ref{eq:decay2})) to the
decay width. The needed modifications to compute the continuum part
(Eqs.~(\ref{eq:decay4}) and~(\ref{eq:decay3})) are discussed after
Eq.~(\ref{eq:nor}).}
\begin{eqnarray}        
\Gamma^{\,(\alpha)} & = & C^{\,(\alpha)} \sum_{N=nljm >
F}\frac12\sum_{m_{s_\Lambda}} \int \frac{d^{\,3} q}{(2 \pi)^{3}}
\frac{1}{2 \omega (q)} 2 \pi \delta (E_{\Lambda}  -
\omega (q) - E_{N}) \nonumber\\ & \times & (G m_\pi^{2})^{2} \left
\vert \left. \left \langle \Lambda , m_{s_\Lambda} \left | \left [ S -
\frac{P}{m_\pi} \vec{\sigma}\cdot \vec{\nabla}_\pi \right ] \right |
nljm; \widetilde{\varphi}_{\pi}^{(\alpha)} (\vec{q})^{\ast}
\right.  \right \rangle \right \vert^2 \label{eq:lambda-decay} 
\end{eqnarray}
where $nljm$ and $E_{N}$
stands for the quantum numbers and relativistic energy 
 of the outgoing nucleon in
the $nlj$ shell, $| \Lambda , m_{s_\Lambda}\rangle$ denotes the
$\Lambda$ wave function with angular momentum third component
$m_{s_\Lambda}$, $E_{\Lambda}$ the $\Lambda$ energy, including its mass,
 $\omega (q)$ the pion energy, and the sum over $N$
runs over the unoccupied nuclear orbitals  ($n$, $l$, $j$,
$m$).  In Eq.~(\ref{eq:lambda-decay}) the sums over $N$ are over proton
or neutron orbitals according to $\alpha$. The pion wave function
$(\widetilde{\varphi}_{\pi}^{(\alpha)} (\vec{q}, x)^{\ast})$ as a
block corresponds to an incoming solution of the Klein Gordon
equation,
\begin{equation}
\left [- \vec{\bigtriangledown}^{2} + m_\pi^{2} + 2 \omega(q) V_{\rm opt}
(\vec{x})\right ] 
\widetilde{\varphi}_{\pi}^{(\alpha)} (\vec{q}, \vec{x})^{\ast} =
(\omega(q) - V_{C} (\vec{x}))^{2} \widetilde{\varphi}_{\pi}^{(\alpha)}
(\vec{q}, \vec{x})^{\ast},
\end{equation}

\noindent
with $V_{C} (\vec{x})$ the Coulomb potential created by the nucleus
considering finite size and vacuum polarization effects, for $\pi^-$
and zero for $\pi^0$, and $V_{\rm opt} (\vec{x})$ the optical
potential which describes the $\pi$-nucleus interaction.  This
potential has been developed microscopically and it is exposed in
detail in Refs.~\cite{NOG93,NOG93bis}. It contains the ordinary lowest
order optical potential pieces constructed from the $s$-- and
$p$--wave $\pi N$ amplitudes. In addition second order terms in both
$s$-- and $p$--waves, responsible for pion absorption, are also
considered.  Standard corrections, as second-order Pauli re-scattering
term, ATT term, Lorentz--Lorenz effect and long and short range
nuclear correlations, are also taken into account. This theoretical
potential reproduces fairly well the data of pionic atoms (binding
energies and strong absorption widths)~\cite{NOG93} and low energy
$\pi$--nucleus scattering~\cite{NOG93bis}.

After a little Racah algebra, one gets for  the $\Lambda$
decay width inside of a nucleus 

\begin{eqnarray}
\Gamma^{(\alpha)} &=& \sum_{N=nlj > F}
\Gamma^{(\alpha)}_{N}\label{eq:1} \\ 
\Gamma^{(\alpha)}_{N} &=& \frac{C^{(\alpha)}}{4 \pi} (G m_\pi^{2})^{2} 
 \frac{q_{N}}{1 + \omega (q_{N}) / E_{A}}
 \left [ S^{2} S^{(s)}_{N} (q_{N}) + \left( \frac{P}{m_\pi} \right)^{2}
\vec{q\,}_{N}^{2} S^{(p)}_{N} (q_{N}) \right ] 
\end{eqnarray}
with
\begin{equation}
q_{N} = ((E_{\Lambda} - E_{N})^{2} - m_\pi^{2})^{1/2}
\end{equation}
Note that $E_N,q_N$ and the integrals $S^{(s)}_N$ and $S^{(p)}_N$,
defined below, depend on the isospin $\alpha$. We
have implemented the recoil factor $(1 + \omega / E_{A})^{-1}$, being
$E_A$ the energy (including the mass) of the daughter nucleus, because
most of the decay corresponds to nucleons in nuclear bound excited
states, and as a consequence the nucleus of mass $M_A$ recoils as a
whole. $S^{(s)}_{N} (q_{N})$ and $S^{(p)}_{N} (q_{N})$ are the
$s$--wave and $p$--wave contributions given by
\begin{eqnarray}
S^{(s)}_{nlj}(q_{N})& =& \frac{2j + 1}{2} \left \vert I_{nlj} (q_N) \right \vert^{2}
\\[2mm]
S^{(p)}_{nlj}(q_{N})& =&
 \frac{1}{\vec{q_N}^{2}} \left \{ l \left \vert M_{nlj} (q_{N}) \right
\vert^{2}~\delta_{j,l-1/2}
+ (l + 1) \vert N_{nlj} (q_{N}) \vert^{2} ~\delta_{j,l+1/2} \right\}
\end{eqnarray}
with
\begin{eqnarray}
I_{nlj} (q_N) &=& \int^{\infty}_{0} r^{2} dr R^{(\Lambda)}_{1s} (r)
R^{(\pi)} _{l} (q_{N} ; r) R_{nlj} (r)
\\[2mm]
M_{nlj} (q_{N}) &=& \int^{\infty}_{0} r^{2} dr R^{(\Lambda)}_{1s}
(r) \left( \frac{d R^{(\pi)} _{l-1} (q_{N} ; r)}{dr} - (l - 1)
\frac{R^{(\pi)} _{l-1} (q_{N} ; r)}{r} \right) \; R_{nlj} (r)
\\[2mm]
N_{nlj} (q_{N}) &=& \int^{\infty}_{0} r^{2} dr R^{(\Lambda)}_{1s} (r)
\left( \frac{d R^{(\pi)} _{l+1} (q_{N} ; r)}{dr} + (l+2)
\frac{R^{(\pi)} _{l+1} (q_{N} ; r)}{r} \right) \; R_{nlj} (r ) .
\end{eqnarray}
Here $R^{(\pi)} _{l} (q_{N} ; r)$ are the radial wave functions of
the pion for each partial wave, regular in the origin and with the
asymptotic behavior
\begin{equation}
R^{(\pi)} _l (q ; r)_{r \rightarrow \infty} \simeq e^{i \delta_{l}}
\frac{1}{qr} \; \sin (qr - l \frac{\pi}{2} + \delta_{l}) \; \; \;
{\rm for} \; \; \; \pi^{0} \label{eq:pi1}
\end{equation}
\begin{equation}
R^{(\pi)} _{l} (q ; r)_{r \rightarrow \infty} \simeq e^{i
(\delta_{l} + \sigma_{l})} \frac{1}{qr} \sin (qr - l \frac{\pi}{2} +
\sigma_{l} + \delta_{l} - \eta \; \log 2 qr) \;\; {\rm for} \; \pi^{-}
\label{eq:pi2}
\end{equation}
with $\eta$ and $\sigma_{l}$ the Coulomb parameter and phase shift 
defined as in Ref.~\cite{GP91} and
$\delta_{l}$ the complex (to take into account inelasticities) phase
shifts obtained from the numerical solution of the Klein Gordon
equation. Finally, $R^{(\Lambda)}_{1s}$ and $R_{nlj} (r)$ are the
$\Lambda$ and nucleon bound radial wave functions, normalized as usual
\begin{equation}
\int_0^{+\infty} dr r^2 |R(r)|^2 = 1 \label{eq:nor}
\end{equation}
The $\Lambda$ wave function in the initial
hypernucleus, $R^{(\Lambda)}$, is obtained from a WS
potential~\cite{Ca99} to account for the mean $\Lambda$--nuclear core
interaction, with parameters compiled in Table~\ref{tab:lam}. The
nucleon dynamics will be studied in detail in  Sect.~\ref{sec:ncs}.

Some of the nuclei which we use are not closed shell nuclei.  In this
case the nucleons from the $\Lambda$ decay can fill up $n_{h}$ empty
states in a $n,l,j$ shell. We take that into account by
multiplying $S^{(s)}_{N}$ and $S^{(p)}_{N}$ by $n_{h}/(2j+1)$.
\begin{table}[tbh]
\begin{center}
\begin{tabular}{ccccc}\hline\\[-3mm]
Hypernucleus & $V_0^\Lambda$ [MeV] & $R^\Lambda$ [fm] & $a^\Lambda$
 [fm] & $B_\Lambda$ [MeV] \\[1mm] \hline
 $^{12}_{\phantom{a}\Lambda}{\rm C}$  & 31.1 & 2.45 & 0.60 & $-10.8$~\cite{Dl88} \\[1mm]
 $^{28}_{\phantom{a}\Lambda}{\rm Si}$ & 30.1 & 3.30 & 0.60 & $-16.6$~\cite{Ha96} \\[1mm]
 $^{56}_{\phantom{a}\Lambda}{\rm Fe}$ & 30.3 & 4.18 & 0.60 & $-21.0$~\cite{Ba84} \\[1mm]
 \hline
\end{tabular}
\end{center}
\caption{\footnotesize Lambda hyperon WS parameters and ground state 
($1s_{\frac12}$) binding energies ($B_\Lambda$). The total $\Lambda$ energy,
$E_\Lambda$ is given by $M_\Lambda+B_\Lambda$.} 
\label{tab:lam}
\end{table}

The above equations~(\ref{eq:lambda-decay})--(\ref{eq:nor}) can
be readily used to compute the discrete contribution to the pionic
decay width, but when the outgoing nucleon coming from the weak
$\Lambda$ decay goes to the continuum (positive energy), one should 
replace the sum $\sum_{N=nlj > F}$ in Eq.~(\ref{eq:1}) by a 
sum over multipoles, plus
an integral over the nucleon energies, i.e., 
\begin{eqnarray}
\sum_{N=nlj > F} &\rightarrow& \sum_{lj} \int_{M}^{
E_{\rm max}} dE ~\frac{2M p}{\pi} \label{eq:cont1} \\[2mm]
R_{nlj}(r) &\rightarrow & R_{lj}(p;r) \label{eq:cont}
\end{eqnarray}
with $E$ and $p = \sqrt{E^2-M^2}$ the nucleon energy and momentum,
$E_{\rm max}=E_\Lambda -m_\pi$, the maximum nucleon energy, neglecting
the recoil of the nucleus, 
and $ R_{lj}(p;r)$ a continuum solution of the nucleon
Schr\"odinger equation~\cite{Am96}, with the same potential as that
used for the bound nucleons to compute the discrete contribution to
the decay width.  The nucleon wave function normalization in the
continuum is the same as that given above for the pions
(Eqs.~(\ref{eq:pi1}-\ref{eq:pi2})).

As a test of the multipole expansion in the continuum, we recovered
the free space $\Lambda $--decay width of Eq.~(\ref{eq:free-decay})
from the multipolar expansion of Eqs.~(\ref{eq:1})
to~(\ref{eq:cont}).  For this purpose we set $A=1$ and replaced the
radial pion and nucleon wave functions by spherical Bessel's
functions, $j_l$, and the radial $\Lambda$--wave function by
$\sqrt{4\pi /V}$, being $V=4\pi L^3/3$ the volume of interaction,
which will be finally sent to infinity when calculating physical
observables. The test is straightforward, taking into account:
\begin{eqnarray}
&&\kern-2cm \sum_{l=0}^\infty (2l+1) \left | \sqrt {\frac{4\pi}{V}} \int_0^L dr
r^2 j_l (pr) j_l(qr) \right |^2  \nonumber \\
&=& \frac{4\pi}{V} \int_0^L dr r^2 \left ( \sum_{l=0}^\infty (2l+1)
j_l^2(pr) \right ) \int_0^L dr r^2
j_l (pr) j_l(qr)  + {\cal O} (1/L) \nonumber \\
&=& \frac{\pi}{2p^2}\delta(p-q) + {\cal O} (1/L) \label{eq:delta}
\end{eqnarray}
On the other hand, we have also used  this free space limit to test
our computational code and the precision  of our numerical algorithms
to solve differential equations, to perform integrations and the sum
over an infinite number of multipoles. The idea
was, taking as starting point the formulae of 
Eqs.~(\ref{eq:1}) to~(\ref{eq:cont}) with a radial
$\Lambda$-- wave function given by $\sqrt{4\pi /V}$ and switching off
the pion-nucleus, and the nucleon
potentials, to recover numerically, in the limit $L$ going to infinity,
the free space $\Lambda$ decay width. Such a calculation, from the
numerical point of view, is much more demanding that the actual 
calculation of the pionic decay width of the hypernuclei, since the
Dirac's delta appearing in Eq.~(\ref{eq:delta}) has to be constructed
numerically out of a large sum over multipoles and slowly
convergent integrals.

\section{Nuclear Structure and Energy Balance }
\label{sec:ncs}
We model the nuclear  structure of the $^{A-1} Z $ system by a
Slater determinant built with single--particle wave functions obtained
by diagonalizing a WS potential well
\begin{equation}\label{eq:pot}
   V_{\rm WS} (r)  
=  V_0 f(r,R_0,a_0) +  V_{LS}
   \frac{\vec {l}\cdot \vec{\sigma}}{r}
   \frac{df(r,R_{LS},a_{LS})}{dr}+ 
    {\tilde V}_C(r),
\end{equation}
where
\begin{equation}
f(r,R,a) =  \frac{1}{1+\exp \left(\frac{r-R}{a} \right)}
\end{equation}
and ${\tilde V}_C(r)$ is the Coulomb potential created by an homogeneous charge
distribution of radius $R_C$. The parameters, compiled in
Table~\ref{tab:nucn} for neutrons and Table~\ref{tab:nucp} for protons, of
this potential are adjusted\footnote{Essentially, we adjust the depths 
$V_0$ and $V_{LS}$, and for the radius-- and thickness--type
parameters standard values have been used} to reproduce some experimental
single--particle energies around the Fermi
level~\cite{Ama94}-\cite{Ama97}. This will enforce not only a correct
energy balance for the decay process to the first available shell but
 also to some excited states. The main contributions to 
the processes of Eqs.~(\ref{eq:decay1}) and~(\ref{eq:decay2}) come
from situations where the daughter nucleus is left in the ground state
or in the first few excited states. Since the effect of the Pauli
blocking depends strongly on the pion energy after the decay, it
is important, as shown in Ref.~\cite{NO92}, to perform a 
correct balance of energies, using when possible experimental
energies. 

Thus, the energy of the first
non-occupied shell is fixed to the mass difference between the ground
states of the $^A Z$ and $^{A-1} Z$ nuclei for the case of neutron
($\pi^0$) decay and of the $^A (Z+1)$ and $^{A-1} Z$ for the case of
the proton ($\pi^-$) decay channel.
\begin{table}[t]
\begin{center}
\begin{tabular}{ccccccc}\hline\\[-3mm]
   & $V_0$ [MeV] & $R_0$ [fm] & $a_0$ [fm] & $V_{LS}$ [MeV] &$R_{LS}$
 [fm] & $a_{LS}$ [fm] \\[1mm]  \hline\\[-3mm]
$^{11}{\rm C}$  & $-$57.0  & 2.86  & 0.53  & $-$6.05  & 2.86 & 0.53 \\[1mm] 
$^{27}{\rm Si}$ & $-$66.2  & 3.50  & 0.70  & $-$3.30  & 3.75 & 0.70  \\[1mm] 
$^{55}{\rm Fe}$ & $-$54.0  & 4.70  & 0.50  & $-$8.30  & 4.70  & 0.50  \\[1mm]  \hline 
\end{tabular}
\end{center}
\caption{\footnotesize Neutron WS parameters used in this work.} 
\label{tab:nucn}
\end{table}
\begin{table}[t]
\begin{center}
\begin{tabular}{cccccccc}\hline\\[-3mm]
   & $V_0$ [MeV] & $R_0$ [fm] & $a_0$ [fm] & $V_{LS}$ [MeV] &$R_{LS}$
 [fm] & $a_{LS}$ [fm] & $R_C$ [fm]\\[1mm]  \hline\\[-3mm]
$^{11}{\rm C}$  & $-$38.4 & 2.86 & 0.53 & $-$6.05 & 2.86 & 0.53 & 2.86 \\[1mm] 
$^{27}{\rm Si}$ & $-$47.4 & 3.75 & 0.53 & $-$10.0 & 3.75 & 0.53 & 3.75 \\[1mm] 
$^{55}{\rm Fe}$ & $-$50.8 & 4.70 & 0.50  & $-$8.30  & 4.70  & 0.50  & 4.70  \\[1mm]  \hline 
\end{tabular}
\end{center}
\caption{\footnotesize Proton WS parameters used in this work.} 
\label{tab:nucp}
\end{table}
In Table~\ref{tab:fermilevel} we give the first available shells and
their empirical energies obtained in this way. 
The energies of this table do not totally
determine the parameters of the mean field 
nucleon potential,  and one has still the
possibility to fit some  excited state energies. In what follows, 
we give some more details of the adjusted shells. 
\begin{table}[tbh]
\begin{center}
\begin{tabular}{c|cc|cc}\hline\\[-3mm]
Hypernucleus & Neutron Shell & Energy [MeV] & Proton Shell& Energy
[MeV] \\[1mm] \hline
 $^{12}_{\phantom{a}\Lambda}{\rm C}$  &$1p_{3/2}$ &$-18.72$ &$1p_{1/2}
$&$-0.60$\\[1mm]
 $^{28}_{\phantom{a}\Lambda}{\rm Si}$ &$1d_{5/2}$ &$-17.18$ &$2s_{1/2}
$&$-2.07$\\[1mm] 
 $^{56}_{\phantom{a}\Lambda}{\rm Fe}$ &$2p_{3/2}$ &$-11.20$ &$1f_{7/2}
$&$-5.85$\\[1mm]  
 \hline
\end{tabular}
\end{center}
\caption{\footnotesize Single particle energies for the first available, to
the pionic decay of the hypernuclei  studied in this work, nucleon shells and
their empirical binding energies. Energies have been obtained from the neutron 
and proton separation energies of the nuclear $^A Z$ and $^A (Z+1)$
spices respectively, and have been taken from Ref.~\cite{Fi96}. }
\label{tab:fermilevel}
\end{table}
\newpage
\begin{itemize}
\item $^{12}_{\phantom{a}\Lambda}{\rm C}$:  
We assume a closed proton and neutron $1p_{3/2}$
shell configurations for the ground state of $^{12}$C and a $1p_{1/2}$
(proton) $1p_{3/2}^{-1}$ (neutron) configuration for the ground state
of $^{12}$N. The rest of neutron and proton shells adjusted by the 
mean field potentials, which parameters are given in  Tables~\ref{tab:nucn}
and~\ref{tab:nucp}, are:
\paragraph {\it Neutrons:}   We fix the energy of the $1p_{1/2}$
neutron shell 
from the energy of the first excited state of $^{12}$C [$2^+$
$^{12}$C$^*$(4.43) MeV].  We assume a particle-hole configuration
$1p_{1/2},1p_{3/2}^{-1}$ which leads to a binding energy for the
$1p_{1/2}$ shell of $-14.29$ MeV. We are aware that this excited
state of $^{12}$C might also have some contributions from proton
degrees of freedom, configuration
mixing, $2p2h$ or higher excitations,  
collective degrees of freedom, etc,   not considered in
this simple model. However, as long as this state had a sizeable
component $|1p_{1/2},1p_{3/2}^{-1}\rangle $, it would be reachable in
the weak decay of the hypernucleus and the procedure described above
would guarantee that the energy balance is correct, not only when the 
$^{12}$C is left on its ground state, but also when it is left on its 
first excited state, leading then in both cases to 
good pion wave--functions. The nuclear matrix elements appearing in the
evaluation of the decay width, though depending strongly on the pion
wave--function, are less sensitive to the specific details of the
nuclear wave function.

\paragraph {\it Protons:} There are no $^{12}$N excited states
amenable to be explained within a shell model, since all levels compiled
in Ref.~\cite{Fi96} are broad resonant states populated in nuclear
reactions. Therefore, we choose to use the same
spin-orbit force as in the neutron case, and 
fix the depth of the central part of the potential to reproduce the
$1p_{1/2}$ shell as given in Table~\ref{tab:fermilevel}. The mean field
potential  adjusted in this way, does not provide
excited states for $^{12}$N.

\item $^{28}_{\phantom{a}\Lambda}{\rm Si}$: We assume  closed proton
and neutron $1d_{5/2}$ shell configurations for the ground state of
$^{28}$Si and a $2s_{1/2}$ (proton) $1d_{5/2}^{-1}$ (neutron)
configuration for the ground state of $^{28}$P. The rest of neutron
and proton shells adjusted by the mean field potentials are:

\paragraph {\it Neutrons:} We fix the energy of the $2s_{1/2}$ neutron
shell from the energy of the first excited state of $^{28}$Si [$2^+$
$^{28}$Si$^*$(1.78) MeV].  We assume a neutron particle-hole
configuration of the type $|2s_{1/2},1d_{5/2}^{-1}\rangle $ which
leads to a binding energy for the $2s_{1/2}$ shell of $15.40$
MeV. Limitations and virtues of describing the mesonic decay with this
simple picture for the underlying nuclear core, are similar to those commented
above for the $^{12}_{\phantom{a}\Lambda}{\rm C}$ case. We should
mention that to adjust the WS neutron potential to simultaneously give
the empirical $1d_{5/2}$ and $2s_{1/2}$ energy shells, keeping the
$1d_{3/2}$ energy shell above the $2s_{1/2}$, is delicate and that we
had to use a value a bit high (0.70 fm) for the thickness parameters
$a_0$ and $a_{LS}$. We have also tried smaller values for the
thickness. For instance for values of $a_0=a_{LS}=0.58$ fm, in order
to adjust the empirical energies of both the $1d_{5/2}$ and $2s_{1/2}$
shells, a spin orbit force very small ($V_{LS}=-0.1$ MeV with
$V_0=-65.75$ MeV) is required. Despite that both the $1d_{5/2}$ and
$1d_{3/2}$ shells turn out to be almost degenerate, and that the latter
one is now deeper than the $2s_{1/2}$ shell, the decay width is rather
stable, and it gets increased only by about 10\%, with respect to the
results of the next section. This increase is due
to an enhancement of the $1d_{3/2}$ shell contribution, but the
corresponding configuration $|1d_{3/2},1d_{5/2}^{-1}\rangle $ has an
excitation energy too small to 
correspond to any experimental excited state of $^{28}$Si.

\paragraph {\it Protons:} For the  $^{28}$P nucleus,  as for the case of
$^{12}$N, there are no excited states amenable to be explained within
a shell model, since all levels compiled in Ref.~\cite{Fi96} are again
broad
resonant states populated in nuclear reactions. We fix the depth of
the central part of the potential to reproduce the $2s_{1/2}$ shell as
given in Table~\ref{tab:fermilevel} and we choose the depth of the
spin-orbit such that the next shell ($1d_{3/2}$) appears in the
continuum.  We will discuss in the next section, 
the dependence of our results on the
precise value of the spin-orbit force. If we use instead the same spin-orbit
force as in the neutron case, there will appear excited $^{28}$P states,
which cannot be identified in the experiment.

\item $^{56}_{\phantom{a}\Lambda}{\rm Fe}$: The nuclear core structure
of this hypernucleus is more difficult to describe, within our simple
shell model, than those of $^{28}_{\phantom{a}\Lambda}{\rm Si}$ and 
$^{12}_{\phantom{a}\Lambda}{\rm C}$ hypernuclei. Thus, our results for
the decay of this hypernucleus are
subject to more theoretical uncertainties. In what the ground
states respect, for $^{56}$Fe, we assume a configuration composed of 
two paired $1f_{7/2}$ proton holes and two paired $2p_{3/2}$
neutron particles, while for $^{56}$Co, we assume a proton hole  in the
$1f_{7/2}$ shell and a neutron in the $2p_{3/2}$ shell.

The rest of neutron
and proton shells adjusted by the mean field potentials are:

\paragraph {\it Neutrons:} We fix the energy of the $1f_{5/2}$ neutron
shell from the
energy of the first excited state of $^{56}$Fe [$2^+$ $^{56}$Fe$^*$(0.847)
MeV].  We assume for  neutrons a two particle configuration
$|2p_{3/2},1f_{5/2}\rangle $ which leads to a binding energy for the
$1f_{5/2}$ shell of $10.35$ MeV.

\paragraph {\it Protons:} The first excited state of $^{56}$Co has
spin-parity $3^+$ and an excitation energy of 0.16 MeV. In principle,
one might use it to determine properties of the WS proton
mean field potential. However, a word of caution must be said here. 
One might try to describe this state as a proton
configuration with two paired holes in the $1f_{7/2}$ shell and a
particle in the $2p_{3/2}$ shell, in such a way that the above
configuration would determine the energy of the $2p_{3/2}$
shell.  The $1f_{7/2}$
shell completes 28 protons which is a magic number, and thus one
expects an energy gap between this shell and the following in
energy, $2p_{3/2}$, appreciable and of the order of the MeV, and not
as small as 0.16 MeV. Then, likely, the $3^+$ excited state should
have an important neutron component (one neutron in the $2p_{3/2}$
shell and the other in the $1f_{5/2}$ one) or more complex 
components   not
considered in our simple shell model.   Then, it seems safe to
guarantee that this state will not have  a sizeable component
$|2p_{3/2},1f_{7/2}^{-2}\rangle $.  Therefore, we choose to use the same
spin-orbit force as in the neutron case and 
fix the depth of the central part of  the potential to reproduce the
$1f_{7/2}$ shell as given in Table~\ref{tab:fermilevel}. The mean field
potential  adjusted in this way, leads to an excited state of about 
5 MeV above the ground state.

\end{itemize}

\section{Results and Concluding Remarks}
\label{sec:con}
 Results for the pionic decay widths of
$^{12}_{\phantom{a}\Lambda}{\rm C}$, $^{28}_{\phantom{a}\Lambda}{\rm
Si}$ and $^{56}_{\phantom{a}\Lambda}{\rm Fe}$, calculated with two
different pion nucleus optical potentials are presented in
Table~\ref{tab:tab1}. The effect of the imaginary part of the potential is
to remove from the emerging pion flux those pions which undergo quasielastic
scattering or pion absorption. However, while the pions absorbed should
be definitely removed, this is not the case with those which undergo
quasielastic scattering, since even if they collide, they are still
there and will be observed. This means that one should not remove
these pions from the pion flux and we take this into account here. The
effect is moderately small, as it was already noted in
Ref.~\cite{NO92}. 
\begin{table}
\begin{center}
\makebox[0cm]{\begin{tabular}{cc|cccc|cccc}\hline\tstrut
& & \multicolumn{4}{c|}{$\pi^0$ Decay} &  \multicolumn{4}{c}{$\pi^-$
Decay}  \\\hline\tstrut 
 $^{A}_{\Lambda}{\rm Z}$ & $V_{\rm opt}$ & $\Gamma_d$ & $ \Gamma_c$ &
$\Gamma$ & $\Gamma_{\rm exp}$~\protect\cite{Sa89}  & $\Gamma_d$ & $ \Gamma_c$ &
$\Gamma$ & $\Gamma_{\rm exp}$~\protect\cite{Sa01}  \\\hline\tstrut
 & FP & 0.136 & 0.008 & 0.144 &
& 0.079 & 0.027 & 0.106 & \\[-2mm]
$^{12}_{\phantom{a}\Lambda}{\rm C}$        & & & & & $0.217 \pm
0.084$ & & & & $0.113 \pm 0.013\pm0.005$ \\[-2mm]
        & NQ & 0.150 & 0.008 & 0.158 & 
& 0.082 & 0.028 & 0.110 & \\\hline\tstrut
& FP & 0.061 & 0.001 & 0.062 &
& 0.018 & 0.019 & 0.037 & \\[-2mm]
$^{28}_{\phantom{a}\Lambda}{\rm Si}$       & & & & & & & & & $0.047 \pm
0.008\pm0.002$ \\[-2mm]
        & NQ & 0.074 & 0.001 & 0.075 &
& 0.020 & 0.019 & 0.039 & \\\hline\tstrut
& FP    & 0.010 & 0.003 & 0.013  &
& 0.005 & 0.009 & 0.014 & \\[-2mm]
$^{56}_{\phantom{a}\Lambda}{\rm Fe}$       & & & & & & & & & $ < 0.015~
(90\% {\rm C.L.}) $ \\[-2mm]
& NQ    & 0.010 & 0.003 & 0.013 &
& 0.004 & 0.010 & 0.014  &\\\hline
\end{tabular}}
\end{center}
\caption{ \footnotesize Pionic decay widths, units of
$\Gamma_\Lambda$, calculated with two different pion nucleus optical
potentials: FP stands for the full optical potential of
Ref.~\protect\cite{NOG93bis}, NQ stands for the pion-nucleus
interaction obtained
by switching off the imaginary part of the FP optical potential 
coming from quasielastic pion scattering.}
\label{tab:tab1}
\end{table}

The agreement with the recent KEK measurements is remarkably good and
it is also good when our results are compared to the older measurement
of the $\pi^0$ decay width of $^{12}_{\phantom{a}\Lambda}{\rm
C}$. This is a clear success of the model of Refs.~\cite{NOG93,
NOG93bis} to account for the pion-nucleus dynamics at low energies. In
Ref.~\cite{NO92}, it was obtained a value of 0.086 $\Gamma_\Lambda$
for the $\pi^-$-$^{12}_{\phantom{a}\Lambda}{\rm C}$ decay width. In
this work we find a value about 25\% higher and in a closer agreement
with the experiment. Differences are even bigger if one looks at the
continuum contribution and also appear for the $\pi^0$ decay case.  In
both works, here and in that of Ref.~\cite{NO92}, the same
$\pi^-$-wave function has been used, being then the difference due to
an improved treatment of the underlying nuclear core dynamics. As we
will see below, the barrier method employed in Ref.~\cite{NO92} to
estimate the continuum contribution compare reasonably well to the
more correct treatment followed here, when the same nuclear potential
is used. The
discrepancies have to be attributed not only to the different WS
potentials used in both works, but also to the somewhat artificial
procedure followed in Ref.~\cite{NO92} to enforce the correct energy
balance in the decay.
\begin{figure}[htb]
\vspace{-3.5cm}
\begin{center}                                                               
%\leavevmode
\epsfysize = 800pt
\makebox[0cm]{\epsfbox{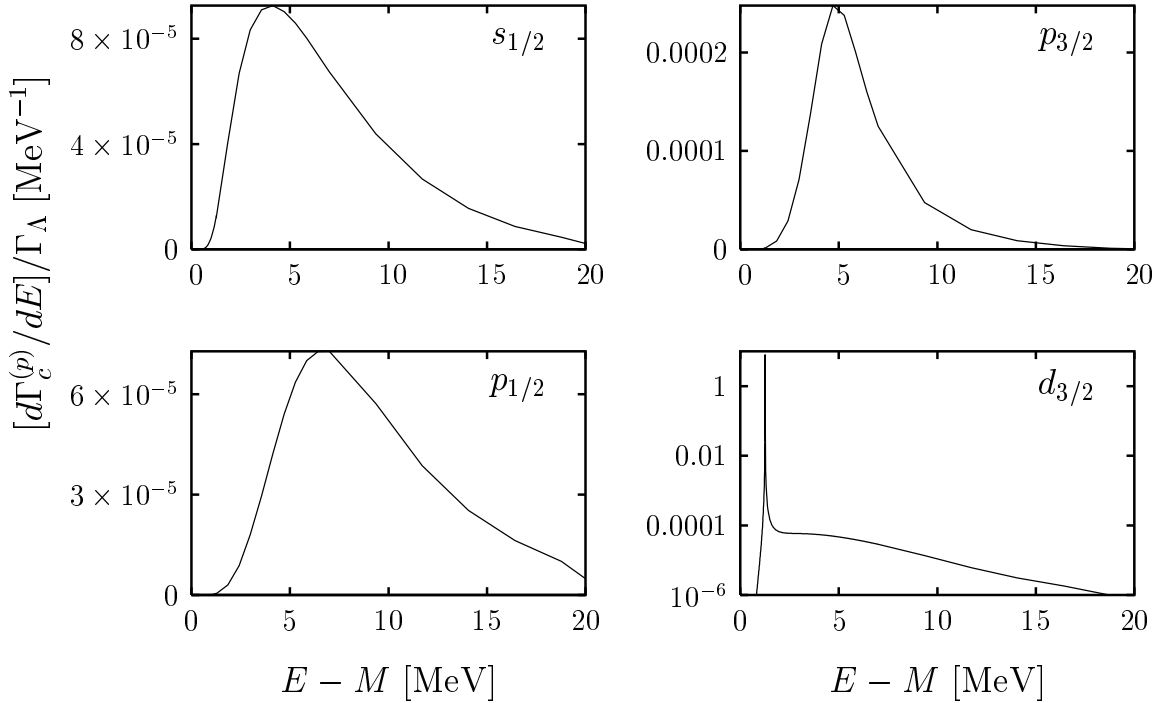}}
\end{center}
\vspace{-15.5cm}
\caption[pepe]{\footnotesize Continuum proton energy distributions
from the $\pi^-$- decay
width of $^{28}_{\protect\phantom{a}\Lambda}{\rm Si}$. 
Results have been obtained with the NQ 
$\pi^-$-nucleus optical potential, as defined in the caption of
Table~\protect\ref{tab:tab1}, and only the most relevant multipoles are shown.}
\label{fig:fig1}
\end{figure}
\begin{figure}[ht]
\vspace{-2.5cm}
\begin{center}                                                               
%\leavevmode
\epsfysize = 600pt
\makebox[0cm]{\epsfbox{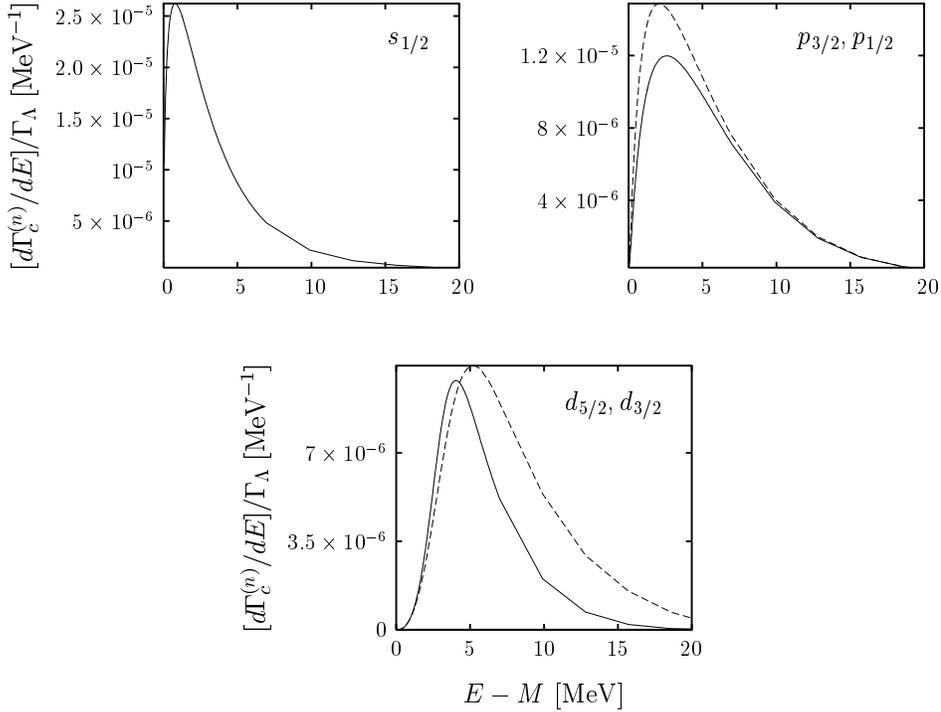}}
\end{center}
\vspace{-9.5cm}
\caption[pepe]{\footnotesize Neutron energy distributions 
for $^{28}_{\protect\phantom{a}\Lambda}{\rm Si}$-decay. Solid (dashed)
line corresponds to the $j=l+1/2$ ($j=l-1/2$) multipole. The
integrated decay widths 
are $1.2 \times 10^{-4}$ , $1.1 \times 10^{-4}$ , $1.0 \times 10^{-4}$ , 
$9 \times 10^{-5}$, and $6 \times 10^{-5}$ ,  
in units of $\Gamma_\Lambda$, for the $1s_{1/2}$,
$1p_{1/2}$, $1p_{3/2}$, $1d_{3/2}$ and $1d_{5/2}$
multipoles respectively.}
\label{fig:fig2}
\end{figure}
\begin{figure}[p]
\vspace{-2.5cm}
\begin{center}                                                               
%\leavevmode
\epsfysize = 600pt
\makebox[0cm]{\epsfbox{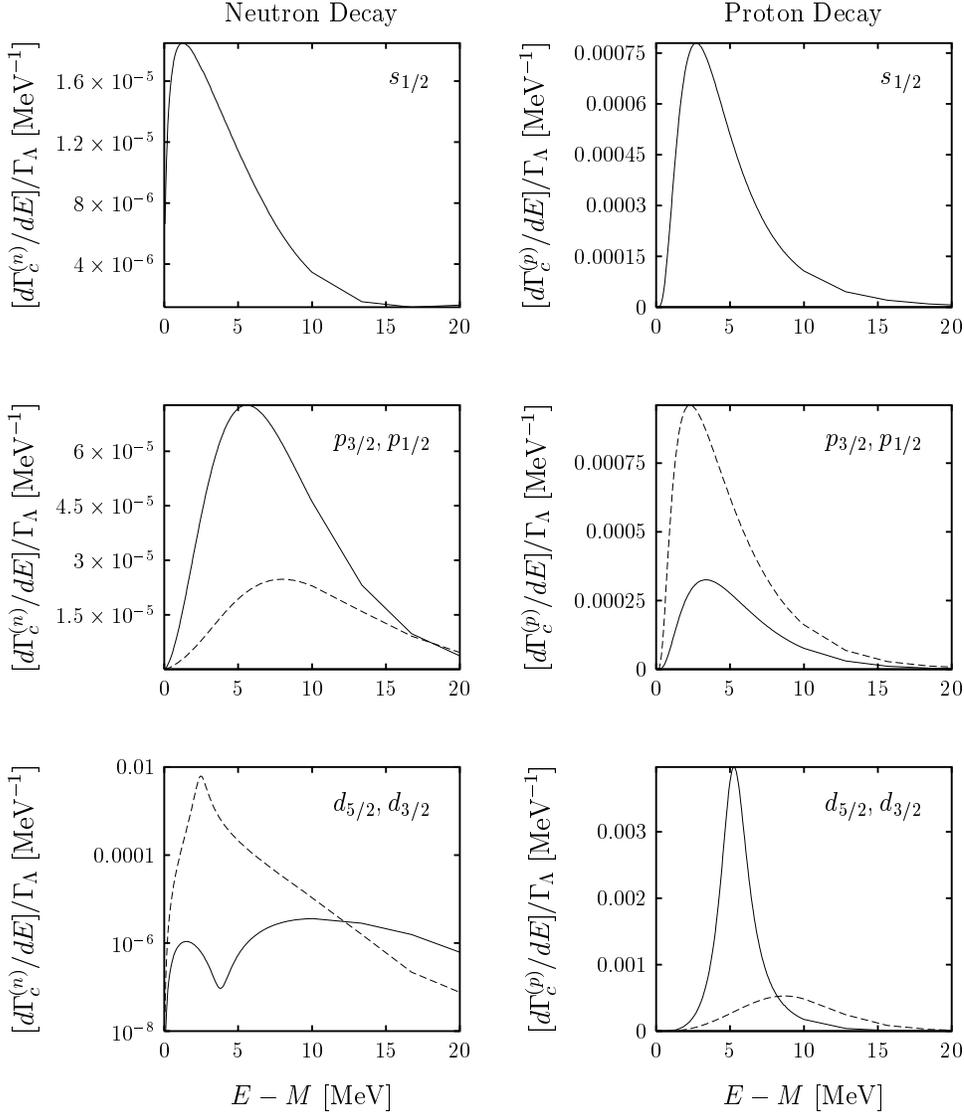}}
\end{center}
\vspace{-4.5cm}
\caption[pepe]{\footnotesize Neutron (left) and proton (right) 
  energy distribution 
for $^{12}_{\protect\phantom{a}\Lambda}{\rm C}$-decay. Solid (dashed)
line corresponds to the $j=l+1/2$ ($j=l-1/2$) multipole. The neutron
 integrated decay widths 
are $1.4 \times 10^{-4}$ , $3.0 \times 10^{-4}$ , $7.1\times 10^{-4}$ , 
$6.62 \times 10^{-3}$, and $ 4 \times 10^{-5}$, 
in units of $\Gamma_\Lambda$, for the $1s_{1/2}$,
$1p_{1/2}$, $1p_{3/2}$, $1d_{3/2}$ and $1d_{5/2}$
multipoles respectively. While for protons,
the contribution of the multipoles 
are $4.33\times 10^{-3}$ , $5.79\times 10^{-3}$ , $2.16\times 10^{-3}$ , 
$4.30 \times 10^{-3}$, and $ 1.124 \times 10^{-2}$. }
\label{fig:fig3}
\end{figure}
\begin{figure}[p]
\vspace{-2.5cm}
\begin{center}                                                               
%\leavevmode
\epsfysize = 600pt
\makebox[0cm]{\epsfbox{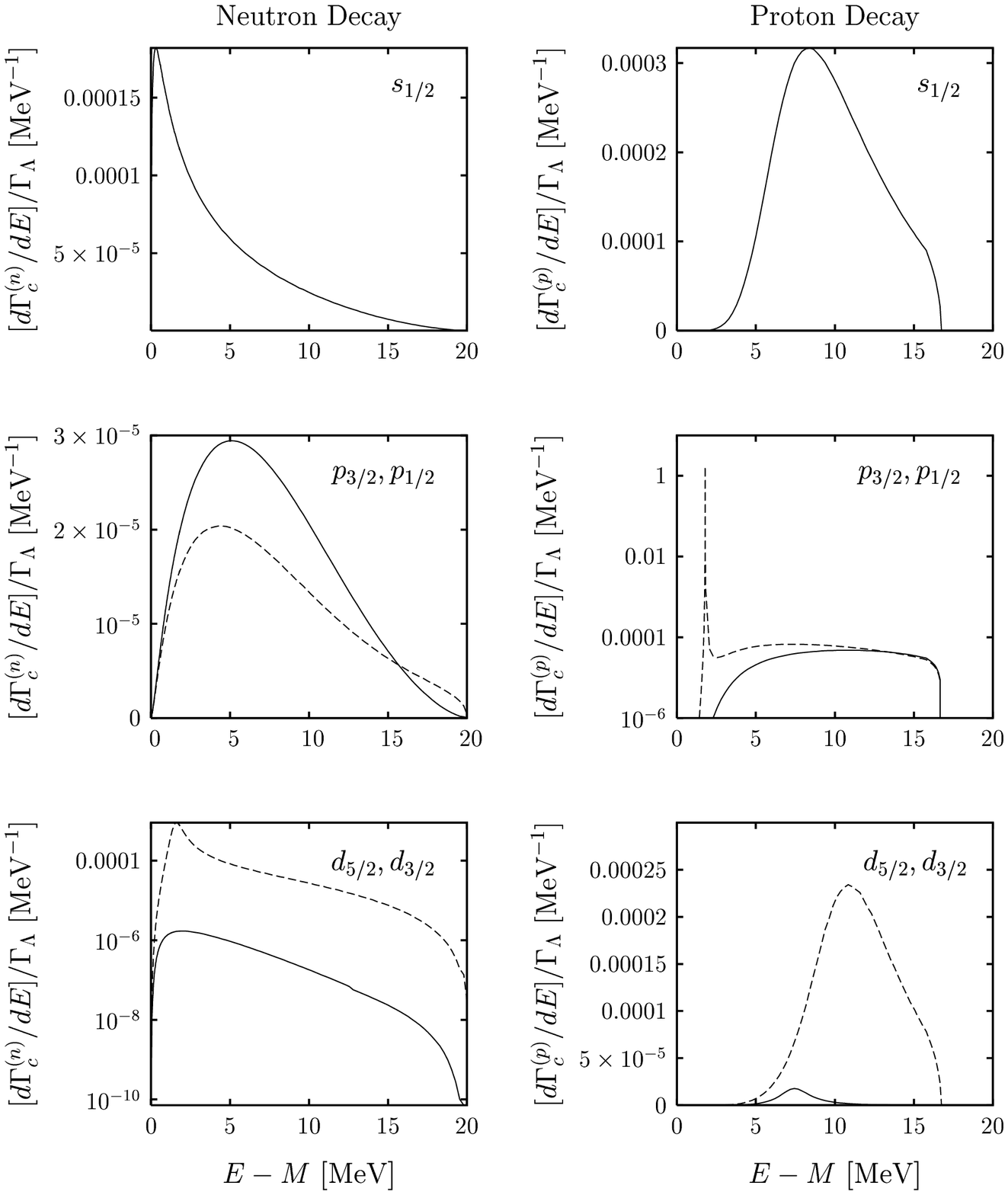}}
\end{center}
\vspace{-4.5cm}
\caption[pepe]{\footnotesize The same as Fig.~\protect\ref{fig:fig3}
for $^{56}_{\protect\phantom{a}\Lambda}{\rm Fe}$-decay.  The neutron
integrated decay widths are $8.2 \times 10^{-4}$ , $2.3 \times
10^{-4}$ , $3.2\times 10^{-4}$ , $1.56 \times 10^{-3}$, and $ 1 \times
10^{-5}$ , in units of $\Gamma_\Lambda$, for the $1s_{1/2}$,
$1p_{1/2}$, $1p_{3/2}$, $1d_{3/2}$ and $1d_{5/2}$ multipoles
respectively. While for protons, the contribution of the multipoles 
are $2.47\times 10^{-3}$, $5.21\times 10^{-3}$ , $4.8\times 10^{-4}$ , $1.49 \times 10^{-3}$,
and $ 5 \times 10^{-5}$.}
\label{fig:fig4}
\end{figure}

Note that in Table~\ref{tab:tab1} the computed widths are always slightly
below the data. The contribution of other reaction channels, not
included in Eqs.~(\ref{eq:decay1}--\ref{eq:decay3}), though we
expect it to be small, would
improve the present calculation. 

In contrast
to most of the hypernuclei studied previously by using the
WFM~\cite{IMB88}--\cite{St93}, for the three hypernuclei considered
here, the continuum contribution in some cases plays a crucial
role. As a matter of example, for $\pi^-$ decay of
$^{28}_{\phantom{a}\Lambda}{\rm Si}$, it turns out that $\Gamma_c$ is
of the same size as $\Gamma_d$ (0.020 vs 0.019) and essential to
understand the experimental datum. This reinforces the need of
updating the calculation of Ref.~\cite{NO92}, where the evaluation of
the continuum contribution was a bit rough, since there was 
assumed that it was
only a small fraction of the total. Indeed, in Ref.~\cite{NO92},
 the continuum contribution was estimated by
discretizing the positive nucleon energy levels, by means of an
infinite barrier placed at distances of about 20 fm. In
Table~\ref{tab:tab2} we compare the CSM of this work with the model of 
Ref.~\cite{NO92}. In both cases we use the same nucleon WS mean
potential.   Both methods agree remarkably well and much better
of what one might expect a priory. Within the model of
Ref.~\cite{NO92}, the sizeable contribution of the continuum is due to
the presence of a quasi-bound (1.27 MeV) state, $1d_{3/2}$ and it has
its counterpart in the size of the $d_{3/2}$ multipole in the
CSM. Even more, the differential partial width $d\Gamma^{\,(p)}_{d_{3/2}} /
dE$ presents a narrow peak (resonance), which gives most of the contribution to
the integrated partial width, located around 1.27 MeV, as can be
seen in Fig~\ref{fig:fig1}.  Small changes in the WS proton mean
potential can bind this shell, going then its important contribution
to the discrete part, $\Gamma_d$, but remaining the total width
$\Gamma_d+ \Gamma_c$ almost unchanged. For example, if one uses a
spin-orbit force depth of $-7$ MeV instead of $-10$ MeV, the $1d_{3/2}$
proton shell becomes bound ($-$0.1 MeV) and  the  total width is 
0.042 $\Gamma_\Lambda$ instead of the value of 0.039 $\Gamma_\Lambda$ quoted in
Table~\ref{tab:tab1}.  In the model of
Ref.~\cite{NO92}, the exact position of the barrier might influence
energies, the number of shells, and contributions of each shell, but
again the total contribution to $\Gamma_c$ remains rather constant, as
long as the barrier is placed far enough.
\begin{table}
\begin{center}
\begin{tabular}{cc|ccc}\hline
\multicolumn{2}{c|}{This work} &
\multicolumn{3}{c}{Barrier at 20 fm~\protect\cite{NO92}} \\ 
Multipole & $\left [\Gamma^{\,(p)}_{lj}\right]_c$ & Shell & Energy [MeV] & 
$\left [\Gamma^{\,(p)}_{nlj}\right ]_c $ \\\hline\tstrut
          &        & $3s_{1/2}$ & 2.45 & 0.0002 \\
          &        & $4s_{1/2}$ & 5.19 & 0.0003 \\
          &        & $5s_{1/2}$ & 9.21 & 0.0002 \\
$s_{1/2}$ & 0.0007& Total $s_{1/2}$ & & 0.0007 \\\hline\tstrut
          &        & $3p_{3/2}$ & 4.59 & 0.0005 \\
          &        & $4p_{3/2}$ & 7.25 & 0.0004 \\
$p_{3/2}$ & 0.0011& Total $p_{3/2}$ & & 0.0009 \\\hline\tstrut
          &        & $3p_{1/2}$ & 5.02& 0.0002 \\
          &        & $4p_{1/2}$ & 8.14    & 0.0002 \\
$p_{1/2}$ & 0.0007& Total $p_{1/2}$ &  & 0.0004 \\\hline\tstrut
          &        & $1d_{3/2}$ & 1.27 & 0.0160 \\
          &        & $2d_{3/2}$ & 3.22 & 0.0001 \\
          &        & $3d_{3/2}$ & 6.31 & 0.0001 \\
$d_{3/2}$ & 0.0165 & Total $p_{1/2}$ & & 0.0162 \\\hline\hline
\multicolumn{2}{c|}{\tstrut $\Gamma_c = 0.019$} & \multicolumn{3}{c}{$\Gamma_c
= 0.018$}\\\hline
\end{tabular}
\end{center}
\caption{\footnotesize Continuum contributions to the pionic decay
width, units of $\Gamma_\Lambda$, calculated with the NQ
$\pi^-$-nucleus optical potential, as defined in the caption of
Table~\protect\ref{tab:tab1}, and two different methods: the sum of
multipoles of the type $\Gamma^{\,(p)}_{lj}$ defined in
Eqs.~(\protect\ref{eq:1}) and~(\protect\ref{eq:cont1}) and the sum
over positive energy discrete bound (by the effect of an
infinite barrier placed at 20 fm) states ~\protect\cite{NO92}. Only
contributions to $\Gamma_c$ larger than $5 \times 10^{-4} ~\Gamma_\Lambda$ are
shown. Results are for the $\pi^-$-decay of the 
$^{28}_{\protect\phantom{a}\Lambda}{\rm
Si}$ hypernucleus.}
\label{tab:tab2}
\end{table}

We have tested for the sensitivity of our results to the mean field
nucleon WS parameters. Thus, we have increased and decreased the spin-orbit
depth $V_{LS}$ by  10\% and re-adjusted the depth of the central part of
the nucleon potential, $V_0$, to get the experimental ground states
masses of the involved nuclei, it is to say $V_0$ is modified
to reproduce again the energies given in Table~\ref{tab:fermilevel},
with the new values of the spin-orbit force. Results turn out to be
quite stable, changing at most by a 2\%, except for the $\pi^-$- decay of 
$^{56}_{\protect\phantom{a}\Lambda}{\rm Fe}$, where the uncertainty
can be as large as one part in fifteen.

To finish, we show  the 
  continuum nucleon energy distributions for the first nucleon
  multipoles, which give the bulk of the total, 
(Figs.~\ref{fig:fig1} to ~\ref{fig:fig4})  and the
contribution to $\Gamma_d$ of each of the unoccupied shells 
 (Table~\ref{tab:discr})  
for  the $\pi^-$ and $\pi^0$ decay widths of
$^{12}_{\protect\phantom{a}\Lambda}{\rm C}$, of
$^{28}_{\protect\phantom{a}\Lambda}{\rm Si}$ and of
$^{56}_{\protect\phantom{a}\Lambda}{\rm Fe}$.  All results have been
obtained using the NQ $\pi$-nucleus optical potential. Resonances
appear as  distinctive features in the continuum contribution of 
some multipoles.
\begin{table}
\begin{center}
\begin{tabular}{c|ccl|ccl}\hline
 & \multicolumn{3}{c|}{$\pi^0$ Decay} & \multicolumn{3}{c}{$\pi^-$ Decay}\\\hline\tstrut 
 $^{A}_{\Lambda}{\rm Z}$ & Shell & Energy [MeV] & $\left
 [\Gamma_{nlj}^{(n)}\right]_d$ & Shell & Energy [MeV] & $\left
 [\Gamma_{nlj}^{(p)}\right]_d$ \\\hline\tstrut 
& $1p_{3/2}$ & $-18.72$ & 0.0473 & & &  \\
& $1p_{1/2}$ & $-14.29$ & 0.0865 & $1p_{1/2}$ & $-0.60$ & 0.0817  \\
& $1d_{5/2}$ & ~$-3.51$  & 0.0156 & & & \\
& $2s_{1/2}$ & ~$-2.07$  & 0.0008 & & & \\
$^{12}_{\phantom{a}\Lambda}{\rm C}$ & \multicolumn{2}{c}{Total} & 0.150 
&\multicolumn{2}{c}{Total} & 0.082 \\\hline\tstrut 
& $1d_{5/2}$ & $-17.18$ & 0.0086 & & &  \\
& $2s_{1/2}$ & $-15.40$ & 0.0285 & $2s_{1/2}$ & $-2.07$ & 0.0204  \\
& $1d_{3/2}$ & $-14.76$ & 0.0342 & & & \\
& $2p_{3/2}$ & ~$-3.29$  & 0.0009 & & & \\
& $1f_{7/2}$ & ~$-2.95$  & 0.0010 & & & \\
& $2p_{1/2}$ & ~$-2.35$  & 0.0009 & & & \\
$^{28}_{\phantom{a}\Lambda}{\rm Si}$ & \multicolumn{2}{c}{Total} & 0.074 
&\multicolumn{2}{c}{Total} & 0.020 \\\hline\tstrut 
&            &          &        & $1f_{7/2}$   & $-5.85$  & 0.0013 \\
& $2p_{3/2}$ & $-11.20$ & 0.0029 & $2p_{3/2}$   & $-0.37 $ & 0.0031 \\
& $1f_{5/2}$ & $-10.35$ & 0.0027 & & &  \\
& $2p_{1/2}$ & ~$-8.84$ & 0.0035  & & &  \\
& $1g_{9/2}$ & ~$-6.33$ & 0.0001  & & &  \\
& $2d_{5/2}$ & ~$-1.08$ & 0.0002  & & &  \\
& $3s_{1/2}$ & ~$-0.33$ & 0.0008  & & &  \\
$^{56}_{\phantom{a}\Lambda}{\rm Fe}$ & \multicolumn{2}{c}{Total} & 0.010 
&\multicolumn{2}{c}{Total} & 0.004\\\hline\tstrut 
\end{tabular}
\end{center}
\caption{\footnotesize Discrete contributions to the pionic decay
width (units of $\Gamma_\Lambda$) for each of the unoccupied shells,
 calculated with the NQ
$\pi-$nucleus optical potential, as defined in the caption of
Table~\protect\ref{tab:tab1}.}
\label{tab:discr}
\end{table}

The study of the mesonic decay of $\LL$ hypernuclei constitutes 
an obvious extension of this work. To improve the existing
calculations~\cite{Ca99, IUM01}  would require a correct
treatment of the $\LL$ pair inside of the nuclear medium. Thus, the
recent work of Ref.~\cite{prl}, where short and long range
correlations are taken into account, will be a good starting point for
this end.

\section*{Acknowledgments}
This research was supported by DGI and FEDER funds, under contract
BFM2002-03218 and by the Junta de Andaluc\'\i a (Spain). C. Albertus
wishes to acknowledge a PhD fellowship from Junta de Andaluc\'\i a.

\end{document}